\documentclass[12pt]{iopart}

\usepackage{amssymb}
\usepackage{graphicx}

\begin{document}

\title{Evolution with size in a locally periodic system:
scattering and deterministic maps}

\author{V Dom\'{\i}nguez-Rocha and M Mart\'inez-Mares}
\address{Departamento de F\'isica, Universidad Aut\'onoma
Metropolitana-Iztapalapa, Apartado Postal 55-534, 09340 M\'exico Distrito
Federal, Mexico}
\eads{\mailto{vdr@xanum.uam.mx}, \mailto{moi@xanum.uam.mx}}

\begin{abstract}
In this paper we study the evolution of the wave function with the system size
in a locally periodic structure. In particular we analyse the dependence of the
wave function with the number of unit cells, which also reflects information
about its spatial behaviour in the system. We reduce the problem to a
nonlinear map and find an equivalence of its  energy regions of single
periodicity and of weak chaos, with the forbidden and allowed bands of the fully
periodic system, respectively. At finite size the wave function decays
exponentially with system size, as well as in space, when the energy lies inside
a region of single periodicity, while for energies in the weak chaotic region
never decays. At the transition between those regions the wave function still
decays but in a $q$-exponential form; we found that the decay length is a half
of the mean free path, which is larger than the lattice constant. 
\end{abstract}


\maketitle


\section{Introduction}

Crystalline materials occupy a special place in the solid state physics.
Despite that in real life the majority of the solids are non crystalline,
crystals help to understand a lot of the properties of the solid matter. At the
macroscopic scale crystalline solids are considered as infinite periodic
systems, which are well described by the band theory~\cite{Ashcroft}. However,
the propagation of electrons through a locally periodic system, which consists
of a finite number $N$ of repeating elements~\cite{Griffiths}, has been of great
interest due to the practical applications in designing artificial materials
with specific features. That is the case of layered periodic structures or
finite superlattices~\cite{Pacher,Morozov1,Morozov2}, and man-made devices
fabricated with optical lattices to model condensed matter
systems~\cite{Courtade,Ponomarev,Wang,Olson,Houston}. Furthermore, the
applicability of locally periodic structures cover a wide range that includes
structurally chiral materials~\cite{Adrian}; microwave~\cite{German},
photonic~\cite{Lipson,Estevez,Archuleta}, and
phononic~\cite{Deymier2007,Deymier2008} crystals, as well as elastic or acoustic
systems~\cite{Sigalas,Flores,Dehesa}.

The study of locally periodic structures is mainly concerned with the band
formation for several cases of fixed $N$, where each one is analysed separately
to look for the emergence of the allowed and forbidden
bands~\cite{Kouwenhoven,Sprung,Pereyra,Exner}. In all cases, precursors of the
band structure of the corresponding fully periodic system, which is almost
formed when $N$ is large enough, can be observed~\cite{Griffiths}. An
interesting question that immediately arises is how or when the properties of a
fully periodic systems are reached. Of course, the standard method of band
theory for crystals is not applicable for finite systems. Therefore, to answer
this question it is necessary to study the evolution of the system as $N$
increases. This procedure allows to get additional information about the band
formation; that is, the evolution to the infinite system should not be the same
for energies in the gap than those in an allowed band of the fully periodic
system. 

Along this line, the electronic transport on a double Cayley tree was studied as
a function of the generation, which plays the same role as
$N$~\cite{Moises2009}. There, a remarkable equivalence between the electronic
transport and the dynamics of an intermittent low-dimensional nonlinear map has
been exhibited; the map presents regions of weak chaos and of single
periodicity, whose Lyapunov exponents are zero and negative, which
indicates the conducting and insulating phases, respectively~\cite{Jiang}. The
conductance oscillates with the generation in the weakly chaotic attractors of
the map, indicating conducting states. In the attractors of single periodicity,
the conductance decays exponentially as is typical for insulating states, but at
the transition the intermittency of the map makes the conductance to reach the
insulating phase in a $q$-exponential form. Therefore, there exist typical
length scales on the different energy regions that remain to be understood.

A scaling analysis of the conductance is one way to obtain information about the
metallic or insulating behaviour of a system~\cite{Lee}. But, the conductance is
a global quantity and not much can be said about the local character in the
system. Therefore, it is important to study the wave function directly, whose
spatial behaviour should reflect whether a system behaves as a conductor or
not. 

The purpose of the present paper is to analyse the evolution of the wave
function with $N$ and once this is done, look at its spatial behaviour in a
locally periodic structure in one dimension. This study allows to determine the
nature of the several states in the system, to answer the question posted above,
and to find a physical interpretation of the typical lengths scales in the
problem. In order to do this, we consider a locally periodic potential that is
formed by a sequence of individual scattering potentials. For our purpose, it is
not necessary to consider an open system on both edges. Therefore, without any
loss of generality, we avoid unnecessary complications and restrict the system
to have only one entrance. We solve the problem analytically for arbitrary
individual potentials using the scattering matrix formalism. Following the same
ideas of reference~\cite{Moises2009} we reduce the problem to the analysis of the
dynamics of an intermittent low-dimensional nonlinear map.

We organize the paper as follows. In the next section we present the scattering
approach to our linear chain of scatterers. In the same section, the recursive
relation satisfied by the scattering matrix, when an individual scatterer is
added, is reduced to a nonlinear map. The scaling analysis and the spatial
dependence of the wave function is presented in section~\ref{sec:scaling}. We
present our conclusions in section~\ref{sec:conclusions}.

\section{A serial structure of quantum scatterers}

\subsection{Scattering approach}

\begin{figure}
\begin{center}
\includegraphics[width=8.0cm]{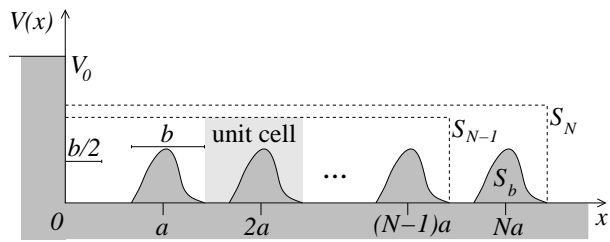}
\end{center}
\caption{A serial structure of identical scatterers, each one represented by a 
potential $V_b$ of arbitrary shape and range $b$, whose associated scattering
matrix is $S_b$. The chain is bounded on the left side by a potential step and
opened on the right side.}
\label{fig:serial}
\end{figure}

The quantum system that we consider is a one-dimensional locally periodic whose
separation between adjacent scatterers is $a$, that we will refer to it as 
the lattice constant. This chain consists of $N$ identical scatterers, each one
described by the potential $V_b(x)$ of arbitrary shape and range $b$, as shown
in figure~\ref{fig:serial}. The unit cell consists of a free region of size
$a-b$ plus the potential of range $b$, except in the first one where the size of
the free region is $a-b/2$. Our chain lies in the semispace at the right of the
origin, bounded on the left ($x=0$) by a potential step of high $V_0\gg E$,
where $E$ is the energy of the quantum particle; the right side of the chain
remains open. We assume that each scatterer is described by a unitary scattering
matrix $S_b$ that has the following structure:
\begin{equation}
\label{eq:Sbarrier}
S_b = 
\left(\begin{array}{cc}
r_b  &  t'_b \\
t_b  &  r'_b
\end{array}\right),
\end{equation}
where $r_b$ ($r'_b$) and $t_b$ ($t'_b$) are the reflection and transmission
amplitudes for incidence on the left (right) of the potential, respectively. 

As we will see below, the evolution of several properties of the system with the
number of scatterers can be obtained from the scattering matrix of the system.
Therefore, we write the scattering matrix $S_N$, that describes the system with
$N$ scatterers, in terms of the one with $N-1$ of them, $S_{N-1}$, by adding an
individual scatterer using the combination rule of scattering matrices. The
combination of $S_{N-1}$ and $S_b$ gives the following recursive relation,
namely
\begin{equation}
\label{eq:Srecursive}
S_N = r'_b + t_b\frac{1}{e^{-2ik(a-b)}-S_{N-1}r_b}S_{N-1}t'_b, 
\end{equation}
where $k=\sqrt{2ME/\hbar^2}$, with $M$ the mass of the particle. The initial
condition for this recursive relation is the scattering matrix $S_0$ associated
to the scattering due to the potential step but measured at $x=b/2$.
 
\subsection{Reduction to a nonlinear map}

Since $S_N$ relates the amplitude of the outgoing plane wave to the amplitude
of the incoming one to the system, it is a $1\times 1$ unitary matrix that can
be parameterized just by a phase, $\theta_N$, as 
\begin{equation}
\label{eq:SN}
S_N = e^{i\theta_N}.
\end{equation}
The recursive relation (\ref{eq:Srecursive}) can be seen as a one-dimensional
nonlinear map. That is, $\theta_N=f(\theta_{N-1})$, where 
\begin{equation}
\label{eq:Phase_recursive}
f(\theta_{N-1}) = - \theta_{N-1} + 
2\arctan \frac{\textrm{Im}\left(
r_b'\alpha_b^*+\alpha_be^{i\theta_{N-1}}\right)}
{\textrm{Re} \left( r_b'\alpha_b^*+\alpha_be^{i\theta_{N-1}}\right)}, 
\end{equation}
modulo $\pi$. Here, $\alpha_b=t_be^{i\phi/2}e^{ik(a-b)}$, $e^{i\phi}=t'_b/t_b$,
and $\textrm{Re}(\alpha')$ and $\textrm{Im}(\alpha')$ denote the real and
imaginary parts of $\alpha'$. Because the reflected wave at the potential step
acquires an additional phase $\theta_{\rm{step}}$, then $S_0=e^{i\theta_0}$
where $\theta_0=\theta_{\rm{step}}+kb$, which is the initial condition for the
map. 

\begin{figure}
\begin{center}
\includegraphics[width=8.0cm]{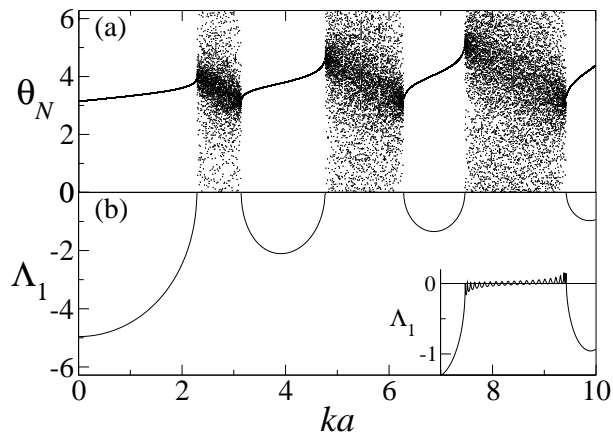}
\end{center}
\caption{(a) Bifurcation diagram for a chain of delta potentials with $ua=10$.
We plot only the last thirty iterations of $N=1000$ for the phase as a
function of $ka$ starting with an initial condition $\theta_0=\pi$. We also
plot the analytical results given by Eqs.~(\ref{eq:fixed-point-delta}) and
(\ref{eq:line}) but they are indistinguishable from the numerics. (b) Finite
$N$ Lyapunov exponent $\Lambda_1(N)$ as a function of $ka$ for $N=1000$.
Theoretical result $\lambda_1$ given by (\ref{eq:lambda1}) is also
plotted; it is indistinguishable from $\Lambda_1(N)$ for very large $N$. Inset:
$\Lambda_1$ for $N=20$.}
\label{fig:bifurcation}
\end{figure}

Although the nonlinear map (\ref{eq:Phase_recursive}) is valid for arbitrary
individual potentials, it is necessary to consider a particular case to look for
the characteristics of the map. Let us assume for a moment that an individual
potential is just a delta potential of intensity $v$ and null range, $b=0$. In
this case, $r_b=r_b'=-u/(u-2ik)$ and $t_b=t_b'=-2ik/(u-2ik)$, where
$u=2Mv/\hbar^2$~\cite{MelloKumar}. The corresponding bifurcation diagram is
shown in figure~\ref{fig:bifurcation}~(a), from which we observe ergodic windows
between windows of periodicity one. The width of an ergodic window depends on
the potential intensity, as well as on the energy; it becomes wider when $ka$
increases (when referring to the delta potential we will speak of $ka$ instead
of $k$), as it is expected since for higher energies, smaller the effect of the
potential. In the ergodic window $\theta_N$ fluctuates, while in the window of
periodicity one it reaches a fixed value when $N$ is very large. 

\begin{figure}
\begin{center}
\includegraphics[width=8.0cm]{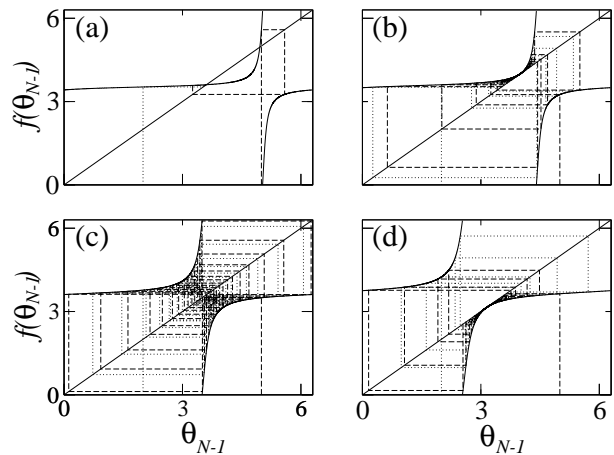}
\end{center}
\caption{Dynamics of the map (\ref{eq:Phase_recursive}) applied to the delta
potential, where $ua=10$ and $ka$ taking the values (a) 2, (b) 2.2846,
(c) 2.8, and (d) 3.14145. Each panel shows the trajectories for initial
conditions $\theta_0=2$ and 5.}
\label{fig:atractor}
\end{figure}

Figure~\ref{fig:bifurcation}~(a) suggests that, in a window of periodicity one,
(\ref{eq:Srecursive}) reaches a fixed point solution for $S_N$ of the form shown
in (\ref{eq:SN}), for very large $N$. If we look for it, we find a stable
and an unstable fixed point solutions, $e^{i\theta_{\pm}}$ [see
figure~\ref{fig:atractor}~(a)]. In an ergodic window, (\ref{eq:Srecursive})
does not have solution of  the form (\ref{eq:SN}), but of the form
$w_{\pm}=|w_{\pm}|e^{i\theta}$, being $\theta$ the value around which $\theta_N$
fluctuates with an invariant density. We will see below that these fixed point
solutions are marginally stable~\cite{Wolf}. The stable and marginally stable
fixed point solutions can be summarized as 
\begin{equation}
\label{eq:Sinfinity}
S_{\infty} = \left\{ 
\begin{array}{clr}
e^{i\theta_{+}} & \mbox{for} & k_{c'_{m-1}} < k < k_{c_m} \\
w_{\pm} & \mbox{for} & k_{c_m} < k < k_{c'_m} \\
e^{i\theta_{-}} & \mbox{for} & k_{c'_m} < k < k_{c_{m+1}}
\end{array}
\right. ,
\end{equation}
where $k_{c_m}$ and $k_{c'_m}$ denote the critical values of $k$ on the left and
right edges of the corresponding ergodic window [see (\ref{eq:edges})
below];
\begin{equation}
\label{eq:fixed-point}
e^{i\theta_{\pm}} = \frac{1}{{r'_b}^*\alpha_b} \left[ 
\pm \sqrt{(\textrm{Re}\,\alpha_b)^2 - |t_b|^4} + i\, \textrm{Im}\,\alpha_b 
\right] 
\end{equation}
for $|\textrm{Re}\,\alpha_b(k)|>|t_b(k)|^2$ and 
\begin{equation}
\label{eq:fixed-w}
w_{\pm} = \frac{i}{{r'_b}^*\alpha_b} \left[ 
\pm \sqrt{|t_b|^4 - (\textrm{Re}\,\alpha_b)^2 } + \textrm{Im}\,\alpha_b 
\right]  
\end{equation}
for 
\begin{equation}
\label{eq:allowed}
|\textrm{Re}\,\alpha_b(k)| \leq |t_b(k)|^2 , 
\end{equation}
such that 
\begin{equation}
\label{eq:theta}
\tan\theta = \frac{\textrm{Im}(ir'_b\alpha_b^*)}{\textrm{Re}(ir'_b\alpha_b^*)}.
\end{equation}
The equality in (\ref{eq:allowed}) determines the critical values
$k_{c_m}$ and $k_{c'_m}$; that is 
\begin{equation}
\label{eq:edges}
|\textrm{Re}\,\alpha_b(k_c)| = |t_b(k_c)|^2 ,
\end{equation}
where $k_c$ denotes $k_{c_m}$ or $k_{c'_m}$. At a critical attractor, $\theta$
takes the value $\theta(k_c)=\theta_c$, where 
\begin{equation}
\tan\theta_{c} = \frac{\textrm{Im}[ir'_b(k_c)\alpha_b^*(k_c)]}
{\textrm{Re}[ir'_b(k_c)\alpha_b^*(k_c)]}.
\end{equation}

The condition (\ref{eq:allowed}) is equivalent to that for allowed bands in the
infinite linear chain of scatterers~\cite{Merzbacher}. Therefore, we have found
a correspondence between windows of single periodicity (or chaotic) and
forbidden (or allowed) bands in the limit $N\to\infty$, in a similar way as
happens for the double Cayley tree studied in Ref.~\cite{Moises2009}. 
Equation~(\ref{eq:edges}) defines the left and right edges of the chaotic
windows. 

The dynamics of the map is expected to be different on each type of windows of
the bifurcation diagram. To observe the trajectories on each region of $k$ we
need to back to the particular case of delta potentials. For this case,
(\ref{eq:fixed-point}) gives 
\begin{equation}
\label{eq:fixed-point-delta}
e^{i\theta_{\pm}} = -i\frac{2ka}{ua}\,e^{-ika} \left[ \pm\, x(ka) + i\, y(ka)
\right] ,
\end{equation}
where
\begin{eqnarray}
x(ka) & = & \sqrt{\left(\cos ka + \frac{ua}{2ka}\sin ka\right)^2 - 1} 
\nonumber \\ 
y(ka) & = & \sin ka - \frac{ua}{2ka}\cos ka 
\end{eqnarray}
for $\left|\cos ka+(ua/2ka)\sin ka\right|>1$, while (\ref{eq:fixed-w})
gives 
\begin{equation}
\label{eq:fixed-w-delta}
w_{\pm} = \frac{2ka}{ua}\, e^{-ika} 
\left[ \pm\, x'(ka) + y(ka) \right] , 
\end{equation}
where 
\begin{equation}
x'(ka) = \sqrt{1-\left(\cos ka + \frac{ua}{2ka}\sin ka\right)^2 }
\end{equation}
for 
\begin{equation}
\label{eq:edges-delta}
\left| \cos ka + \frac{ua}{2ka} \sin ka \right| \leq 1, 
\end{equation}
which is the condition for allowed bands in the Kronig-Penney
model~\cite{Kronig}. From (\ref{eq:fixed-w-delta}) it is easy to see that
the phase of $w_{\pm}$ is given by [see (\ref{eq:theta})] 
\begin{equation}
\label{eq:line}
\theta = -ka + (m+1)\pi, \quad m = 0,\, 1,\ldots 
\end{equation}
The band edges are obtained from the equality in (\ref{eq:edges-delta}),
namely  
\begin{equation}
\tan \frac{k_{c_m}a}2 = 
\left\{ 
\begin{array}{rl}
ua/2k_{c_m}a, & \quad\mbox{for} \, m\, \mbox{odd} \\ 
-2k_{c_m}a/ua, & \quad\mbox{for} \, m\, \mbox{even} 
\end{array}
\right.,
\end{equation}
and $k_{c'_m}a = m\pi$, for $m$ odd and even. At the band edges,
$\theta(k_{c_m}a)=\theta_{c_m}$, with $\theta_{c_m} = -k_{c_m}a + (m+1)\pi$.
For the first chaotic window ($m=1$), $k_{c_1}a=2.28445\ldots$ and
$k_{c'_1}a=\pi$, $\theta_{c_1}=3.99873\ldots$, and $\theta_{c'_1}=\pi$.
Equations~(\ref{eq:fixed-point-delta}) and (\ref{eq:line}) coincide with the
numerical results in both sides and inside of the first chaotic window, as it is
shown in figure~\ref{fig:bifurcation}~(a). 

The trajectories of the map for this example are shown in
figure~\ref{fig:atractor} for some values of $ka$. In panel (a) $ka=2$, $ka$ is
inside the window of period one; we observe that all trajectories (we show only
two) converge to a fixed point as $N\to\infty$, which corresponds to
$\theta_+\approx 3.61$. Panels (b) and (d) show intermittent trajectories
at the tangent bifurcations when $ka=2.2846$ and 3.14145, very close to the
transitions from the chaotic side. In panel (c), deep inside the chaotic
window, the trajectories never converge. Therefore, we observe that the
behaviour of the trajectories for two initial conditions strongly depends on the
specific region to which $ka$ belongs. The dynamics of the map can be analysed
by means of the sensitivity to initial conditions, something that we do next. 

\subsection{Sensitivity to initial conditions}

The dynamics of the nonlinear map (̣\ref{eq:Phase_recursive}) is characterized
by the sensitivity to initial conditions. For finite $N$ it is defined by
\cite{Moises2009}
\begin{equation}
\label{eq:sensitivity}
\Xi_N\equiv\left|{\frac{d\theta_N}{d\theta_0}}\right|\equiv
e^{N\Lambda_1(N)} ,
\end{equation}
where $\theta_0$ is the initial condition and $\Lambda_1(N)$ is the finite $N$ 
Lyapunov exponent, whose dependence on $N$ is shown as an argument (the
subscript 1 will be clear below). For the map (\ref{eq:Phase_recursive}),
$\Xi_N$ satisfies the following recursive relation: 
\begin{equation}
\label{eq:Lambda-iteration}
e^{N\Lambda_1(N)} = 
\frac{\left| t_b \right|^4}
{\left|r_b'\alpha_b^*+\alpha_bS_{N-1}\right|^2} 
e^{(N-1)\Lambda_1(N-1)}.
\end{equation}

In figure~\ref{fig:bifurcation}~(b) we plot $\Lambda_1(N)$ obtained from
(\ref{eq:Lambda-iteration}) for the delta potentials as a function of $ka$,
for $N=1000$. We observe that $\Lambda_1(N)$ is negative in the windows of
period one, indicating that $\Xi_N$ decays exponentially with $N$ when $N$ is
very large. This means that any trajectory in the map converges rapidly to a
fixed point, as happens in figure~\ref{fig:atractor}~(a). In the chaotic
windows, $\Lambda_1(N)$ goes to zero in the limit $N\to\infty$. There, $\Xi_N$
does not depend on $N$ and nothing can be said about the convergence of any
trajectory; this situation corresponds to figure~\ref{fig:atractor}~(c). For a
finite number of iterations $\Lambda_1(N)$ is still negative in the windows of
period one, but it oscillates around zero in the chaotic windows. This can be
seen in the inset of figure~\ref{fig:bifurcation}; the amplitude of those
oscillations tend to zero as $N$ increases; this is a signal of weak
chaos~\cite{Jiang,Sven}. Here, we are interested in the behaviour of $\Xi_N$
with $N$.

\begin{figure}
\begin{center}
\includegraphics[width=8.5cm]{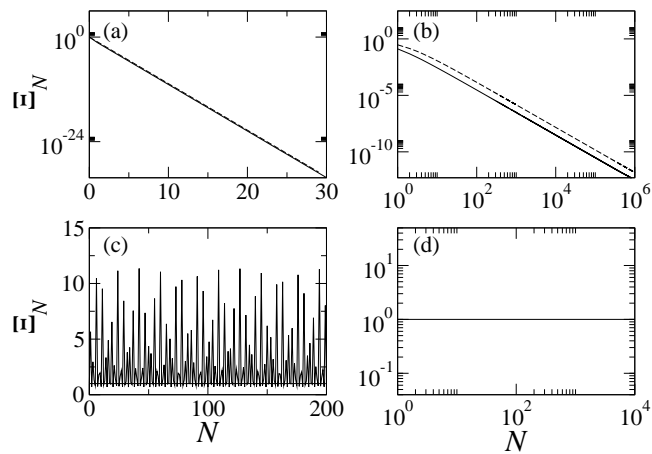}
\end{center}
\caption{The continuous lines represent the result for the sensitivity to
initial conditions for finite $N$, $\Xi_N$, as a function of $N$ for the chain
of delta potentials with $ua=10$ and $ka$ equal to (a) 2, (b) $2.28445\ldots$,
(c) 2.8, and (d) 3.14145. The dashed lines are the theoretical results given by
(\ref{eq:xin}) for (a) and (c), and (\ref{eq:xinq}) for (b). In (d)
$\xi_N$ remains constant with $N$.}
\label{fig:xin}
\end{figure}

With this evidence, we can assume that, in the limit $N\to\infty$,
$\Lambda_1(N)\to\lambda_1$ and $\Xi_N\to\xi_N$, where $\xi_N$ is the sensitivity
to initial conditions defined by 
\begin{equation}
\label{eq:xin}
\xi_N = e^{N\lambda_1}, 
\end{equation}
where $\lambda_1$ is the Lyapunov exponent of the map, which is given by 
\begin{equation}\
\label{eq:lambda1}
\lambda_1 = \ln\frac{|t_b|^4}
{\left|r'_b\alpha_b^*+\alpha_bS_{\infty}\right|^2}.
\end{equation}
For one of the two roots expressed in (\ref{eq:fixed-point}), which are
valid for $k$ in a window of period one, $\lambda_1$ is positive. This means
that the fixed point solution is unstable. The second solution is a stable one
since $\lambda_1$ is negative. It is the last one, the only that appears in
figure~\ref{fig:bifurcation}~(a). In the chaotic windows there are two values of 
$\lambda_1$ that correspond to the two solutions $w_{\pm}$, one positive and
another negative, both at the same distance from zero. We find that the Lyapunov
exponent that agrees with $\Lambda_1(N)$, which has
been obtained iteratively for very large $N$, is the average of those values,
which is zero. In figure~\ref{fig:bifurcation}~(b) we compare $\lambda_1$ of
(\ref{eq:lambda1}) with $\Lambda_1(N)$ calculated iteratively by means of
(\ref{eq:Lambda-iteration}), for the delta potentials. For $N=1000$ we
observe that both results are indistinguishable. Therefore, (\ref{eq:xin})
says that $\xi_N$ decays exponentially with $N$ for $k$ in the windows of period
one and remains constant in the chaotic windows. These results are verified in
panels (a) and (c) of figure~\ref{fig:xin}, where we compare (\ref{eq:xin})
with the numerical calculations for the delta potential. 

The behaviour of the sensitivity to initial conditions is very different at the
critical attractors, since there an anomalous dynamics occurs due to the tangent
bifurcation~\cite{Geisel,Baldovin}. A critical attractor is located at the point
$(k_c,\,\theta_c)$, where $k_c$ denotes $k_{c_m}$ or $k_{c'_m}$ and $\theta_c$,
$\theta_{c_m}$ or $\theta_{c'_m}$. If we make an expansion of $\theta_N$ close
to $\theta_c$ the result is given by 
\begin{equation}
\theta_N - \theta_c = \left( \theta_{N-1} - \theta_c \right) + 
u \left( \theta_{N-1} - \theta_c \right)^z + \cdots, 
\end{equation}
where $z=2$ and $u=\mp |r'_b(k_c)|^2$. From known properties of this
nonlinearity of the tangent bifurcation, the sensitivity obeys a $q$-exponential
law for large $N$~\cite{Baldovin}. That is, $\Xi_{N\to\infty}=\xi_N$, where
\begin{equation}
\label{eq:xinq}
\xi_N \propto e_q^{N\lambda_q} \equiv 
\left[ 1 - (q-1)N\lambda_q \right]^{-1/(q-1)}, 
\end{equation}
with $q=1-1/z=3/2$ and $\lambda_{3/2}=zu=\mp 2|r'_b(k_c)|^2$. The minus and plus
signs correspond to trajectories at the left and right (right and left) of the
point of tangency $\theta_c=\theta_{c_m}$ ($\theta_c=\theta_{c'_m}$),
respectively; that is, $\xi_N$ decays with $N$ with a power law when
$\theta_{N-1}-\theta_{c_m}<0$ ($\theta_{N-1}-\theta_{c'_m}>0$) and grows faster
than exponential when  $\theta_{N-1}-\theta_{c_m}>0$
($\theta_{N-1}-\theta_{c'_m}<0$). 

In panels (b) and (d) of figure~\ref{fig:xin} we show the behaviour of $\Xi_N$
with $N$ at the edges $k_{c_1}a$ and $k_{c'_1}a$, obtained from the numerical
calculation for the delta potential, and compare them with (\ref{eq:xinq}).
In (b) it is clear that $\Xi_N$ behaves as $\xi_N$ of (\ref{eq:xinq}) at
$ka=k_{c_1}a$ for very large $N$. That is not the case at the right edge
$ka=k_{c'_1}a$ where $\Xi_N$ remains constant with $N$. This pathological
behaviour is because $k_{c'_1}a$ corresponds to a resonance and the nodes the
delta
potentials become invisible. 

\section{Behaviour of the wave function}
\label{sec:scaling}

\subsection{Evolution of the wave function with the system size}

The wave function in the region between the individual potentials can be written
as a superposition of plane waves traveling to the left and right. If we
normalize it, in such a way that the amplitude after the last scatterer is one,
the square modulus of the wave function in the region between the scatterers $n$
and $n+1$, labeled by $n$ ($n=0,\,1,\dots,\,N$), can be written as 
\begin{equation}
\label{eq:Psi-spatial}
\left| \psi^{(N)}_n(x) \right|^2 = 
\frac{e_q^{N\Lambda_q(N)}}
{e_q^{n\Lambda_q(n)}} 
\cos^2 \left[ k(x-na-b/2) + \theta_n/2 \right] ,
\end{equation}
where $q=1$ for $k$ in the windows of single period and of weak chaos; $q=3/2$
at the transition from the chaotic side. We see from (\ref{eq:Psi-spatial})
that at the position $x$, the amplitude of the wave function depends on $N$ only
through the numerator. This means that we can replace $N$ by $N-1$ to find the
wave function at the same position for a chain made of $N-1$ scatterers. Hence,
the following recursive relation is satisfied by the square modulus of the wave
function: 
\begin{equation}
\left| \psi^{(N)}_n(x) \right|^2 = 
\frac{e_q^{N\Lambda_q(N)}}
{e_q^{(N-1)\Lambda_q(N-1)}} 
\left| \psi^{(N-1)}_n(x) \right|^2 .
\end{equation}
An iteration process lead us to 
\begin{equation}
\label{eq:PsiNn}
\left| \psi^{(N)}_n(x) \right|^2 = e_q^{N\Lambda_q(N)}
\left| \psi^{(0)}_n(x) \right|^2 ,
\end{equation}
where $\psi^{(0)}_n(x)$ is the wave function at $x$ in the absence of any
scatterer.

The factor in front of the right hand side of (\ref{eq:PsiNn}) is just the
sensitivity to initial conditions for finite $N$ [see (\ref{eq:sensitivity})].
Therefore, the evolution of $|\psi^{(N)}_n(x)|^2$ with $N$ is given by $\Xi_N$,
as shown in figure~\ref{fig:xin}. This evolution resemble the behaviour of the
conductance in a double Cayley tree with system size~\cite{Moises2009}. For
energies in a window of period one, $|\psi^{(N)}_n(x)|^2$ decays exponentially
with $N$, for $N\gg 1$, with a typical decay length which we identify with a
localization length due to the similar scaling behaviour of the conductance with
$N$~\cite{Moises2009}; this localization length is given by 
\begin{equation}
\label{eq:zeta1}
\zeta_1 = \frac{a}{|\lambda_1|}.
\end{equation} 
When $|\lambda_1|>1$, on the one hand, the localization length is smaller than
the lattice constant; this occurs far from the transition to the chaotic side;
close to the transition $\zeta_1>a$. On the other hand, for energies in the
chaotic window, $|\psi^{(N)}_n(x)|^2$ never decays but oscillates as $N$
increases; there, $\lambda_1=0$ such that a localization length can not be
defined. However, at the transition from the chaotic side the wave function
shows a power law decay with $N$ [see figure~\ref{fig:xin}(b)]. The typical
decay length, that we identify with a localization length too, is given by 
\begin{equation}
\label{eq:zeta3/2}
\zeta_{3/2} = \frac{a}{|\lambda_{3/2}|} = 
\frac{a}{2|r'_b(k_c)|^2} .
\end{equation}
What is very interesting here is that the term on the right hand side of the
second equality is $\ell/2$, where 
\begin{equation}
\label{eq:mfp}
\ell = \frac{a}{\left|r'_b(k_c)\right|^2},
\end{equation}
which is a definition of the mean free path~\cite{MelloPRL1988,MelloStone}. This
equation means that the mean free path is larger than the lattice constant. This
result is in agreement with the one obtained from Esaki and Tsu, where $\ell\sim
3a$, in a supperlattice~\cite{Esaki}. For our particular case of a delta
potential with $ua=10$, $\ell\approx 1.21a$ for $k_{c_1}a=2.28445\ldots$, and
$\ell\approx 1.4a$ for $k_{c'_1}a=\pi$. The behaviour of the wave function in
the space can let us to understand the nature of the states on the different
energy regions.

\subsection{Spatial behaviour of the wave function}

\begin{figure}
\begin{center}
\includegraphics[width=8.7cm]{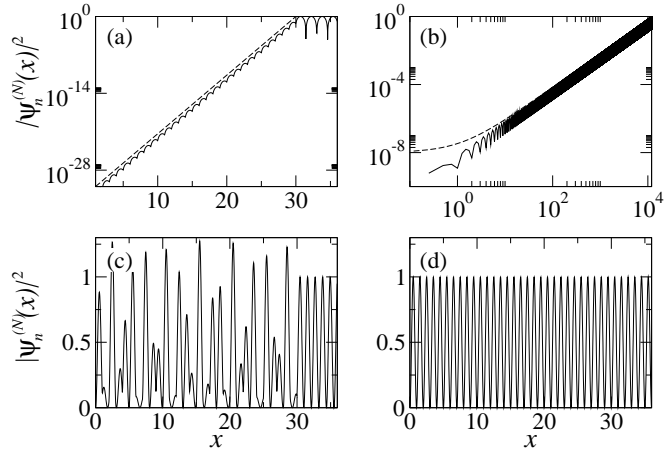}
\end{center}
\caption{Square of the wave function as a function of $x$ for the chain of delta
potentials with $ua=10$. In (a) we compare (\ref{eq:Psi-spatial3})
(continuous line) with an exponential with localization length given by
(\ref{eq:zeta1}) (dashed line), for $ka=2$. In (b) we compare
(\ref{eq:Psi-spatial4}) (continuous line) with a $q$-exponential decay with
localization length given by (\ref{eq:zeta3/2}) (dashed line), for
$ka=2.28445\ldots$. (c) For $ka=2.8$ the wave function is extended. In (c)
the system is invisible at a resonance at $ka=\pi$.}
\label{fig:psin}
\end{figure}

In the limit of very large $N$, $\Lambda_q(N)\to\lambda_q$ and the squared
modulus of the wave function given by (\ref{eq:Psi-spatial}) can be written as 
\begin{equation}
\label{eq:Psi-spatial2}
\left| \psi^{(N)}_n(x) \right|^2 = 
\frac{e_q^{N\lambda_q}}{e_q^{n\lambda_q}} 
\cos^2 \left[ k(x-na-b/2) + \theta_n/2 \right] .
\end{equation}
For $k$ far from the critical attractors, $q=1$ and the ordinary exponential
function is recovered. In this case the spatial behaviour of the wave function
is given by 
\begin{equation}
\label{eq:Psi-spatial3}
\left| \psi^{(N)}_n(x) \right|^2 = 
e^{-(N-n)a/\zeta_1} 
\cos^2 \left[ k(x-na-b/2) + \theta_n/2 \right] .
\end{equation}
This equation means that when the energy is in a window of period one, the wave
function decreases exponentially in space from the boundary ($n=N$) to inside of
the system ($n<N$). In this case the quantum particle is spread over few
$\zeta_1$'s. It is in this sense that we interpret $\zeta_1$ as a localization
length. This localization length is smaller that the period of our locally
periodic structure, $a$, except very close to the transition region. This
situation is illustrated in figure~\ref{fig:psin}~(a) for the chain of delta
potentials.

For energies in the weakly chaotic windows $\zeta_1\to\infty$ and the wave
function becomes extended through the system, as can be seen in
figure~\ref{fig:psin}~(c).

Near the transition, at the weakly chaotic attractors,
(\ref{eq:Psi-spatial2}) can be written as
\begin{equation}
\label{eq:Psi-spatial4}
\left| \psi^{(N)}_n(x) \right|^2 = 
\left(\frac{1\pm na/2\zeta_{3/2}}{1\pm Na/2\zeta_{3/2}}\right)^2 
\cos^2 \left[ k(x-na-b/2) + \theta_n/2 \right] ,
\end{equation}
where we used that $\lambda_{3/2}=\mp a/\zeta_{3/2}$. The plus (minus)
sign in (\ref{eq:Psi-spatial4}) corresponds to the left (right) chaotic
attractor. This equation clearly shows a power law decay from the boundary.
That is, the wave function is still localized but the localization is not
exponential, but $q$-exponential. It is interesting to note that the
localization length is $\zeta_{3/2}=\ell/2$. The factor 1/2 is due to the fact
that the localization length is measured from the maximum of the wave function,
which in this case is at the boundary of the system, while the mean free path
implicitly assumes the width of a wave packet; according to Esaki and
Tsu~\cite{Esaki}, the mean free path in a supperlattice is the uncertainty in
the position. In figure~\ref{fig:psin}~(b) we can observe the power law decay of
the squared modulus of the wave function for the chain of delta potentials.
Figure~\ref{fig:psin}~(d) shows the behaviour of the squared modulus of the wave
function at the attractor on the right side of the chaotic window. In this case,
the system is invisible since $k_{c'_1}a=\pi$ corresponds to a resonance and
the wave function has nodes just at the delta potentials.


\section{Conclusions}
\label{sec:conclusions}

We considered a locally periodic structure in one dimension, which consists of a
chain of potentials of arbitrary shape. We studied the evolution of the wave
function of the chain when the system size increases. This procedure helped us
to observe the behaviour of the wave function in space when the size of the
system remains fixed.

Since the system is not fully periodic, we can not use the traditional band
theory. Instead of that we take advantage of a recursion relation of the
scattering matrix, in terms of the number of scatterers, to reduce the problem
to a nonlinear map. Through this method all information about the behaviour of
the system was obtained from the dynamics of this map. In this way we could
understand how and when the band theory of the fully periodic chain is reached
by the knowledge of the type of evolution of the wave function with the system
size. We found an equivalence between the periodic and weakly chaotic regions of
the map and the forbidden and allowed bands, respectively. This equivalence is
similar to that remarked in the literature for the conductance of a double
Cayley tree, where it is null in the windows of period one, and oscillates in
regions of weak chaos. Also, we found that the wave function at a given position
scales with the system size in a similar way as the conductance does. In a
window of period one, far from the transition to the chaotic window, the wave
function decays exponentially with a typical length scale, which is smaller than
the lattice constant. Very close to the transition from the window of period
one, this typical length becomes larger than the lattice constant. This
behaviour driven us to interpret the typical scale as a localization length. We
corroborated this interpretation by the exponential localization of the wave
function close to the boundary of the finite system. In the chaotic region the
wave function does not decay and a localization length can not be defined.

At the transition between periodic and weakly chaotic regions, but from the
chaotic side, the wave function scales as a power law. This implies that it is
still localized but with a $q$-exponential form. The localization length were
found to be one half of the mean free path. We demonstrate analytically, that
the mean free path is larger than the lattice constant, in complete agreement
with a result found in the literature for a supperlattice. 

Finally, it is worth mentioning that our development considers independent
quantum particles such that classical waves can also be used.  In particular,
one-dimensional elastic systems are strong candidates to simulate our quantum
system by means of elastic rods with narrow notches~\cite{Flores}. 

\ack
The authors thank RA M\'endez-S\'anchez and G~B\'aez useful discussions.
M Mart\'inez-Mares is grateful with the Sistema Nacional de Investigadores,
Mexico, and with MA Torres-Segura for her encouragement. V~Dom\'inguez-Rocha
thanks financial support from CONACyT, Mexico, and partial support from C~Jung
through the CONACyT project No.~79988.

\end{document}